\documentclass[     ,final              ]{aipproc}%
\usepackage{amssymb}
\usepackage{amsmath}
\usepackage{amsfonts}
\usepackage{graphicx}%
\layoutstyle{6x9}
\begin{document}

\title{Effects of Modified Dispersion Relations and Noncommutative Geometry on the
Cosmological Constant Computation}

\author{Remo Garattini}{address={Universit\`{a} degli Studi di Bergamo, Facolt\`{a} di Ingegneria,
\\ Viale Marconi 5, 24044 Dalmine (Bergamo) Italy and\\ I.N.F.N. -
sezione di Milano, Milan, Italy.\\ E-mail:remo.garattini@unibg.it}}

\begin{abstract}
We compute Zero Point Energy in a spherically symmetric background with the
help of the Wheeler-DeWitt equation. This last one is regarded as a
Sturm-Liouville problem with the cosmological constant considered as the
associated eigenvalue. The graviton contribution, at one loop is extracted with
the help of a variational approach together with Gaussian trial functionals.
The divergences handled with a zeta function regularization are compared with
the results obtained using a Noncommutative Geometry (NCG) and Modified
Dispersion Relations (MDR). In both NCG and MDR no renormalization scheme is
necessary to remove infinities in contrast to what happens in conventional approaches.
Effects on photon propagation are briefly discussed.
\end{abstract}
\maketitle

\section{Introduction}

The Cosmological Constant problem is certainly one of the most fascinating
challenges of our days. A challenge because all the attempts that try to
explain the $10^{120}$ orders of magnitude of discrepancy between the theory
and observation have produced unsatisfying results. If we believe that Quantum
Field Theory (QFT) is a part of the real world, then the theoretical
predictive power to compute the Cosmological Constant must be entrusted to the
methods of QFT at the Planck scale. Indeed, calculating the Zero Point Energy
(ZPE) of some field of mass $m$ with a cutoff at the Planck scale, we obtain%
\begin{equation}
E_{ZPE}=\frac{1}{2}\int_{0}^{\Lambda_{p}}\frac{d^{3}k}{\left(  2\pi\right)
^{3}}\sqrt{k^{2}+m^{2}}\simeq\frac{\Lambda_{p}^{4}}{16\pi^{2}}\approx
10^{71}GeV^{4}, \label{zpe}%
\end{equation}
while the observation leads to a ZPE of the order $10^{-47}GeV^{4}$. This
surely represents one of the worst predictions of QFT. However if one insists
to use the methods of QFT applied to General Relativity, one necessarily meets
one of the most famous equations appeared in the literature of Cosmology and
Gravity: the Wheeler-DeWitt (WDW) equation\cite{DeWitt}. The WDW equation was
originally introduced by Bryce DeWitt as an attempt to quantize General
Relativity in a Hamiltonian formulation. It is described by%
\begin{equation}
\mathcal{H}\Psi=\left[  \left(  2\kappa\right)  G_{ijkl}\pi^{ij}\pi^{kl}%
-\frac{\sqrt{g}}{2\kappa}\!{}\!\left(  \,\!^{3}R-2\Lambda\right)  \right]
\Psi=0 \label{WDW}%
\end{equation}
and it represents the quantum version of the classical constraint which
guarantees the invariance under time reparametrization. $G_{ijkl}$ is the
super-metric, $\pi^{ij}$ is the super-momentum,$^{3}R$ is the scalar curvature
in three dimensions and $\Lambda$ is the cosmological constant, while
$\kappa=8\pi G$ with $G$ the Newton's constant. An immediate application of
the WDW Equation is given in terms of the Friedmann-Robertson-Walker (FRW)
mini superspace, where all the degrees of freedom but the scale factor are
frozen. The FRW metric is described by the following line element%
\begin{equation}
ds^{2}=-N^{2}dt^{2}+a^{2}\left(  t\right)  d\Omega_{3}^{2}, \label{FRW}%
\end{equation}
where $d\Omega_{3}^{2}$ is the usual line element on the three sphere, namely%
\begin{equation}
d\Omega_{3}^{2}=\gamma_{ij}dx^{i}dx^{j}.
\end{equation}
In this background, we have simply%
\begin{equation}
R_{ij}=\frac{2}{a^{2}\left(  t\right)  }\gamma_{ij}\qquad\mathrm{and}\qquad
R=\frac{6}{a^{2}\left(  t\right)  }%
\end{equation}
and the WDW equation $H\Psi\left(  a\right)  =0$, becomes%
\begin{equation}
\left[  -a^{-q}\left[  \frac{\partial}{\partial a}a^{q}\frac{\partial
}{\partial a}\right]  +\frac{9\pi^{2}}{4G^{2}}\left(  a^{2}-\frac{\Lambda}%
{3}a^{4}\right)  \right]  \Psi\left(  a\right)  =0. \label{WDW_0}%
\end{equation}
Eq.$\left(  \ref{WDW_0}\right)  $ assumes the familiar form of a
one-dimensional Schr\"{o}dinger equation for a particle moving in the
potential%
\begin{equation}
U\left(  a\right)  =\frac{9\pi^{2}}{4G^{2}}a^{2}\left(  1-\frac{a^{2}}%
{a_{0}^{2}}\right)  \label{U(a)}%
\end{equation}
with total zero energy. The parameter $q$ represents the factor-ordering
ambiguity and $a_{0}=\sqrt{\frac{3}{\Lambda}}$ is a reference length. The WDW
equation $\left(  \ref{WDW_0}\right)  $ has been solved exactly in terms of
Airy functions by Vilenkin\cite{Vilenkin} for the special case of operator
ordering $q=-1$. The Cosmological Constant $\Lambda$ here appears as a
parameter. Nevertheless, except the FRW case and other few examples, the WDW
equation is very difficult to solve. This difficulty increases considerably
when the mini superspace approach is avoided. However some information can be
gained if one changes the point of view. Indeed, instead of treating $\Lambda$
in Eq.$\left(  \ref{WDW}\right)  $ as a parameter, one can formally rewrite
the WDW equation as an expectation value computation\footnote{See also
Ref.\cite{CG:2007} for an application of the method to a $f\left(  R\right)  $
theory.}\cite{Remo}. Indeed, if we multiply Eq.$\left(  \ref{WDW}\right)  $ by
$\Psi^{\ast}\left[  g_{ij}\right]  $ and functionally integrate over the three
spatial metric $g_{ij}$ we find%
\begin{equation}
\frac{1}{V}\frac{\int\mathcal{D}\left[  g_{ij}\right]  \Psi^{\ast}\left[
g_{ij}\right]  \int_{\Sigma}d^{3}x\hat{\Lambda}_{\Sigma}\Psi\left[
g_{ij}\right]  }{\int\mathcal{D}\left[  g_{ij}\right]  \Psi^{\ast}\left[
g_{ij}\right]  \Psi\left[  g_{ij}\right]  }=\frac{1}{V}\frac{\left\langle
\Psi\left\vert \int_{\Sigma}d^{3}x\hat{\Lambda}_{\Sigma}\right\vert
\Psi\right\rangle }{\left\langle \Psi|\Psi\right\rangle }=-\frac{\Lambda
}{\kappa}. \label{VEV}%
\end{equation}
In Eq.$\left(  \ref{VEV}\right)  $ we have also integrated over the
hypersurface $\Sigma$ and we have defined%
\begin{equation}
V=\int_{\Sigma}d^{3}x\sqrt{g}%
\end{equation}
as the volume of the hypersurface $\Sigma$ with%
\begin{equation}
\hat{\Lambda}_{\Sigma}=\left(  2\kappa\right)  G_{ijkl}\pi^{ij}\pi^{kl}%
-\sqrt{g}^{3}R/\left(  2\kappa\right)  . \label{LambdaSigma}%
\end{equation}
In this form, Eq.$\left(  \ref{VEV}\right)  $ can be used to compute ZPE
provided that $\Lambda/\kappa$ be considered as an eigenvalue of $\hat
{\Lambda}_{\Sigma}$. In particular, Eq.$\left(  \ref{VEV}\right)  $ represents
the Sturm-Liouville problem associated with the cosmological constant. To
solve Eq.$\left(  \ref{VEV}\right)  $ is a quite impossible task. Therefore,
we are oriented to use a variational approach with trial wave functionals. The
related boundary conditions are dictated by the choice of the trial wave
functionals which, in our case are of the Gaussian type. Different types of
wave functionals correspond to different boundary conditions. The choice of a
Gaussian wave functional is justified by the fact that ZPE should be described
by a good candidate of the \textquotedblleft\textit{vacuum state}%
\textquotedblright. In the next section we give the general guidelines in
ordinary gravity and in presence of Modified Dispersion Relations and the Non
Commutative approach to QFT. Units in which $\hbar=c=k=1$ are used throughout
the paper.

\section{High Energy Gravity Modification: the Example of Non Commutative
Theories and Gravity's Rainbow}

As an application of the Eq.$\left(  \ref{VEV}\right)  $, we consider the
simple example of the Mini Superspace described by a FRW cosmology $\left(
\ref{FRW}\right)  $. We find the following simple expectation value%
\begin{equation}
\frac{\int\mathcal{D}a\Psi^{\ast}\left(  a\right)  \left[  -\frac{\partial
^{2}}{\partial a^{2}}+\frac{9\pi^{2}}{4G^{2}}a^{2}\right]  \Psi\left(
a\right)  }{\int\mathcal{D}a\Psi^{\ast}\left(  a\right)  \left[  a^{4}\right]
\Psi\left(  a\right)  }=\frac{3\Lambda\pi^{2}}{4G^{2}},\label{WDW_1}%
\end{equation}
where the normalization is modified by a weight factor. The application of a
variational procedure with a trial wave functional of the form%
\begin{equation}
\Psi=\exp\left(  -\beta a^{2}\right)
\end{equation}
shows that there is no real solution of the parameter $\beta$ compatible with
the procedure. Nevertheless, a couple of imaginary solutions of the
variational parameter%
\begin{equation}
\beta=\pm i\frac{3\pi}{4G}%
\end{equation}
can be found. These solutions could be related to the tunneling wave function
of Vilenkin\cite{Vilenkin}. Even if the eigenvalue procedure of Eq.$\left(
\ref{WDW_1}\right)  $ leads to an imaginary cosmological constant, which does
not correspond to the measured observable, it does not mean that the procedure
is useless. Indeed, Eq.$\left(  \ref{VEV}\right)  $ can be used to calculate
the cosmological constant induced by quantum fluctuations of the gravitational
field. To fix ideas, we choose the following form of the metric%
\begin{equation}
ds^{2}=-N^{2}\left(  r\right)  dt^{2}+\frac{dr^{2}}{1-\frac{b\left(  r\right)
}{r}}+r^{2}\left(  d\theta^{2}+\sin^{2}\theta d\phi^{2}\right)  ,\label{dS}%
\end{equation}
where $b\left(  r\right)  $ is subject to the only condition $b\left(
r_{t}\right)  =r_{t}$. We consider $g_{ij}=\bar{g}_{ij}+h_{ij},$where $\bar
{g}_{ij}$ is the background metric and $h_{ij}$ is a quantum fluctuation
around the background. Then we expand Eq.$\left(  \ref{VEV}\right)  $ in terms
of $h_{ij}$. Since the kinetic part of $\hat{\Lambda}_{\Sigma}$ is quadratic
in the momenta, we only need to expand the three-scalar curvature $\int
d^{3}x\sqrt{g}{}^{3}R$ up to second order in $h_{ij}$\footnote{See
Refs.\cite{CG:2007,Remo1:2004,Remo1:2009} for technical details.}. As shown in
Ref.\cite{Remo1:2009}, the final contribution does not include ghosts and
simply becomes%
\begin{equation}
\frac{1}{V}\frac{\left\langle \Psi^{\bot}\left\vert \int_{\Sigma}d^{3}x\left[
\hat{\Lambda}_{\Sigma}^{\bot}\right]  ^{\left(  2\right)  }\right\vert
\Psi^{\bot}\right\rangle }{\left\langle \Psi^{\bot}|\Psi^{\bot}\right\rangle
}=-\frac{\Lambda^{\bot}}{\kappa}.\label{lambda0_2a}%
\end{equation}
The integration over Gaussian Wave functionals leads to%
\begin{equation}
\hat{\Lambda}_{\Sigma}^{\bot}=\frac{1}{4V}\int_{\Sigma}d^{3}x\sqrt{\bar{g}%
}G^{ijkl}\left[  \left(  2\kappa\right)  K^{-1\bot}\left(  x,x\right)
_{ijkl}+\frac{1}{\left(  2\kappa\right)  }\!{}\left(  \tilde{\bigtriangleup
}_{L\!}\right)  _{j}^{a}K^{\bot}\left(  x,x\right)  _{iakl}\right]
,\label{p22}%
\end{equation}
where%
\begin{equation}
\left(  \tilde{\bigtriangleup}_{L\!}\!{}h^{\bot}\right)  _{ij}=\left(
\bigtriangleup_{L\!}\!{}h^{\bot}\right)  _{ij}-4R{}_{i}^{k}\!{}h_{kj}^{\bot
}+\text{ }^{3}R{}\!{}h_{ij}^{\bot}\label{M Lichn}%
\end{equation}
is the modified Lichnerowicz operator and $\bigtriangleup_{L}$is the
Lichnerowicz operator defined by%
\begin{equation}
\left(  \bigtriangleup_{L}h\right)  _{ij}=\bigtriangleup h_{ij}-2R_{ikjl}%
h^{kl}+R_{ik}h_{j}^{k}+R_{jk}h_{i}^{k}\qquad\bigtriangleup=-\nabla^{a}%
\nabla_{a}.\label{DeltaL}%
\end{equation}
$G^{ijkl}$ represents the inverse DeWitt metric and all indices run from one
to three. Note that the term $-4R{}_{i}^{k}\!{}h_{kj}^{\bot}+$ $^{3}R{}%
\!{}h_{ij}^{\bot}$ disappears in four dimensions. The propagator $K^{\bot
}\left(  x,x\right)  _{iakl}$ can be represented as
\begin{equation}
K^{\bot}\left(  \overrightarrow{x},\overrightarrow{y}\right)  _{iakl}%
=\sum_{\tau}\frac{h_{ia}^{\left(  \tau\right)  \bot}\left(  \overrightarrow
{x}\right)  h_{kl}^{\left(  \tau\right)  \bot}\left(  \overrightarrow
{y}\right)  }{2\lambda\left(  \tau\right)  },\label{proptt}%
\end{equation}
where $h_{ia}^{\left(  \tau\right)  \bot}\left(  \overrightarrow{x}\right)  $
are the eigenfunctions of $\tilde{\bigtriangleup}_{L\!}$. $\tau$ denotes a
complete set of indices and $\lambda\left(  \tau\right)  $ are a set of
variational parameters to be determined by the minimization of Eq.$\left(
\ref{p22}\right)  $. The expectation value of $\hat{\Lambda}_{\Sigma}^{\bot}$
is easily obtained by inserting the form of the propagator into Eq.$\left(
\ref{p22}\right)  $ and minimizing with respect to the variational function
$\lambda\left(  \tau\right)  $. Thus the total one loop energy density for TT
tensors becomes%
\begin{equation}
\frac{\Lambda}{8\pi G}=-\frac{1}{2}\sum_{\tau}\left[  \sqrt{\omega_{1}%
^{2}\left(  \tau\right)  }+\sqrt{\omega_{2}^{2}\left(  \tau\right)  }\right]
.\label{1loop}%
\end{equation}
The above expression makes sense only for $\omega_{i}^{2}\left(  \tau\right)
>0$, where $\omega_{i}$ are the eigenvalues of $\tilde{\bigtriangleup}_{L\!}$.
Following Refs.\cite{CG:2007,Remo1:2004,Remo1:2009}, we find that the final
evaluation of expression $\left(  \ref{1loop}\right)  $ is%
\begin{equation}
\frac{\Lambda}{8\pi G}=-\frac{1}{\pi}\sum_{i=1}^{2}\int_{0}^{+\infty}%
\omega_{i}\frac{d\tilde{g}\left(  \omega_{i}\right)  }{d\omega_{i}}d\omega
_{i}=-\frac{1}{4\pi^{2}}\sum_{i=1}^{2}\int_{\sqrt{m_{i}^{2}\left(  r\right)
}}^{+\infty}\omega_{i}^{2}\sqrt{\omega_{i}^{2}-m_{i}^{2}\left(  r\right)
}d\omega_{i},\label{tot1loop}%
\end{equation}
where we have included an additional $4\pi$ coming from the angular
integration and where we have defined two r-dependent effective masses
$m_{1}^{2}\left(  r\right)  $ and $m_{2}^{2}\left(  r\right)  $%
\begin{equation}
\left\{
\begin{array}
[c]{c}%
m_{1}^{2}\left(  r\right)  =\frac{6}{r^{2}}\left(  1-\frac{b\left(  r\right)
}{r}\right)  +\frac{3}{2r^{2}}b^{\prime}\left(  r\right)  -\frac{3}{2r^{3}%
}b\left(  r\right)  \\
\\
m_{2}^{2}\left(  r\right)  =\frac{6}{r^{2}}\left(  1-\frac{b\left(  r\right)
}{r}\right)  +\frac{1}{2r^{2}}b^{\prime}\left(  r\right)  +\frac{3}{2r^{3}%
}b\left(  r\right)
\end{array}
\right.  \quad\left(  r\equiv r\left(  x\right)  \right)  .\label{masses}%
\end{equation}
The effective masses have different expression from case to case. For example,
in the Schwarzschild case, $b\left(  r\right)  =2MG$, we find%
\begin{equation}
\left\{
\begin{array}
[c]{c}%
m_{1}^{2}\left(  r\right)  =\frac{6}{r^{2}}\left(  1-\frac{2MG}{r}\right)
-\frac{3MG}{r^{3}}\\
\\
m_{2}^{2}\left(  r\right)  =\frac{6}{r^{2}}\left(  1-\frac{2MG}{r}\right)
+\frac{3MG}{r^{3}}%
\end{array}
\right.  \quad\left(  r\equiv r\left(  x\right)  \right)  .
\end{equation}
The expression in Eq.$\left(  \ref{tot1loop}\right)  $ is divergent and must
be regularized. For example, the zeta function regularization method leads to%
\begin{equation}
\rho_{i}\left(  \varepsilon\right)  =\frac{m_{i}^{4}\left(  r\right)  }%
{64\pi^{2}}\left[  \frac{1}{\varepsilon}+\ln\left(  \frac{4\mu^{2}}{m_{i}%
^{2}\left(  r\right)  \sqrt{e}}\right)  \right]  \quad i=1,2\quad,\label{rhoe}%
\end{equation}
where an additional mass parameter $\mu$ has been introduced in order to
restore the correct dimension for the regularized quantities. Such an
arbitrary mass scale emerges unavoidably in any regularization scheme. The
renormalization is performed via the absorption of the divergent part into the
re-definition of a bare classical quantity and the final result is given by%
\begin{equation}
\frac{\Lambda_{0}}{8\pi G}=\frac{m_{1}^{4}\left(  r\right)  }{64\pi^{2}}%
\ln\left(  \frac{4\mu^{2}}{m_{1}^{2}\left(  r\right)  \sqrt{e}}\right)
+\frac{m_{2}^{4}\left(  r\right)  }{64\pi^{2}}\ln\left(  \frac{4\mu^{2}}%
{m_{2}^{2}\left(  r\right)  \sqrt{e}}\right)  .\label{lambda0}%
\end{equation}
Of course, one can follow other methods to obtain finite results: for
instance, the use of a UV-cut off. Nevertheless it is possible to obtain
finite results introducing a distortion in the space-time from the beginning.
This can be realized with the help of the Non Commutative Approach to QFT
developed in Ref.\cite{RG PN} or with the help of Gravity's Rainbow developed
in Ref.\cite{RGGM}. Noncommutative theories provide a powerful method to
naturally regularize divergent integrals appearing in QFT. Eq.$\left(
\ref{tot1loop}\right)  $ is a typical example of a divergent integral. The
noncommutativity of spacetime is encoded in the commutator $\left[
\mathbf{x}^{\mu},\mathbf{x}^{\nu}\right]  =i\,\theta^{\mu\nu}$, where
$\theta^{\mu\nu}$ is an antisymmetric matrix which determines the fundamental
discretization of spacetime. In even dimensional space-time, $\theta^{\mu\nu}$
can be brought to a block-diagonal form by a suitable Lorentz rotation
leading to%
\begin{equation}
\theta^{\mu\nu}=diag\left[  \theta_{1}\epsilon^{ab}\theta_{2}\epsilon
^{ab}\ldots\theta_{d/2}\epsilon^{ab}\right]
\end{equation}
with $\epsilon^{ab}$ a $2\times2$ antisymmetric Ricci Levi-Civita tensor. If
$\theta_{i}\equiv\theta$ $\forall i=1\ldots d/2$, then the space-time is
homogeneous and preserves isotropy. The effect of the $\theta$ length on ZPE
calculation is basically the following: the classical Liouville counting
number of nodes%
\begin{equation}
dn=\frac{d^{3}\vec{x}d^{3}\vec{k}}{\left(  2\pi\right)  ^{3}},
\end{equation}
is modified by distorting the counting of nodes in the following way\cite{RG
PN}%
\begin{equation}
dn=\frac{d^{3}xd^{3}k}{\left(  2\pi\right)  ^{3}}\ \Longrightarrow
\ dn_{i}=\frac{d^{3}xd^{3}k}{\left(  2\pi\right)  ^{3}}\exp\left(
-\frac{\theta}{4}\left(  \omega_{i,nl}^{2}-m_{i}^{2}\left(  r\right)  \right)
\right)  ,\quad i=1,2.\label{moddn}%
\end{equation}
This deformation corresponds to an effective cut off on the background
geometry $\left(  \ref{dS}\right)  $. The UV cut off is triggered only by
higher momenta modes $\gtrsim1/\sqrt{\theta}$ which propagate over the
background geometry. As an effect the final induced cosmological constant
becomes%
\[
\frac{\Lambda}{8\pi G}=\frac{1}{6\pi^{2}}\left[  \int_{\sqrt{m_{1}^{2}\left(
r\right)  }}^{+\infty}\sqrt{\left(  \omega^{2}-m_{1}^{2}\left(  r\right)
\right)  ^{3}}e^{-\frac{\theta}{4}\left(  \omega^{2}-m_{1}^{2}\left(
r\right)  \right)  }d\omega\right.
\]%
\begin{equation}
\left.  +\int_{\sqrt{m_{2}^{2}\left(  r\right)  }}^{+\infty}\sqrt{\left(
\omega^{2}-m_{2}^{2}\left(  r\right)  \right)  ^{3}}e^{-\frac{\theta}%
{4}\left(  \omega^{2}-m_{2}^{2}\left(  r\right)  \right)  }d\omega\right]
,\label{t1loop}%
\end{equation}
where an integration by parts in Eq.$\left(  \ref{tot1loop}\right)  $ has been
done. We recover the usual \textit{divergent} integral when $\theta
\rightarrow0$. The result is finite and we have an induced cosmological
constant which is regular. We can obtain enough information in the asymptotic
r\'{e}gimes when the background satisfies the relation
\begin{equation}
m_{0}^{2}\left(  r\right)  =m_{1}^{2}\left(  r\right)  =-m_{2}^{2}\left(
r\right)  ,\label{masses1}%
\end{equation}
which is valid for the Schwarzschild, Schwarzschild-de Sitter (SdS) and
Schwarzschild-Anti de Sitter (SAdS) metric close to the throat. Indeed,
defining%
\begin{equation}
x=\frac{m_{0}^{2}\left(  r\right)  \theta}{4},\label{xS}%
\end{equation}
we find that when $x\rightarrow+\infty$,
\begin{equation}
\frac{\Lambda}{8\pi G}\simeq\frac{1}{6\pi^{2}\theta^{2}}\sqrt{\frac{\pi}{x}%
}\left[  3+\left(  8x^{2}+6x+3\right)  \exp\left(  -x\right)  \right]
\rightarrow0.\label{LNCSz}%
\end{equation}
Conversely, when $x\rightarrow0$, we obtain%
\begin{equation}
\frac{\Lambda}{8\pi G}\simeq\frac{4}{3\pi^{2}\theta^{2}}\left[  2-\left(
\frac{7}{8}+\frac{3}{4}\ln\left(  \frac{x}{4}\right)  +\frac{3}{4}%
\gamma\right)  x^{2}\right]  \rightarrow\frac{8}{3\pi^{2}\theta^{2}}.
\end{equation}
The other interesting cases, namely de Sitter and Anti-de Sitter and Minkowski
are described by%
\begin{equation}
m_{1}^{2}\left(  r\right)  =m_{2}^{2}\left(  r\right)  =m_{0}^{2}\left(
r\right)  ,\label{masses2}%
\end{equation}
leading to%
\begin{equation}
\frac{\Lambda}{8\pi G}\simeq\frac{1}{6\pi^{2}}\left(  \frac{4}{\theta}\right)
^{2}\frac{3}{8}\sqrt{\frac{\pi}{x}}\rightarrow0,
\end{equation}
when $x\rightarrow\infty$ and%
\begin{equation}
\frac{\Lambda}{8\pi G}\simeq\frac{1}{6\pi^{2}}\left(  \frac{4}{\theta}\right)
^{2}\left[  1-\frac{x}{2}+\left(  -\frac{7}{16}-\frac{3}{8}\ln\left(  \frac
{x}{4}\right)  -\frac{3}{8}\gamma\right)  x^{2}\right]  \rightarrow\frac
{8}{3\pi^{2}\theta^{2}},
\end{equation}
when $x\rightarrow0$. As regards Gravity's Rainbow\cite{MagSmo}, we can begin
by defining a\textquotedblleft\textit{rainbow metric}\textquotedblright%
\begin{equation}
ds^{2}=-\frac{N^{2}\left(  r\right)  dt^{2}}{g_{1}^{2}\left(  E/E_{P}\right)
}+\frac{dr^{2}}{\left(  1-\frac{b\left(  r\right)  }{r}\right)  g_{2}%
^{2}\left(  E/E_{P}\right)  }+\frac{r^{2}}{g_{2}^{2}\left(  E/E_{P}\right)
}\left(  d\theta^{2}+\sin^{2}\theta d\phi^{2}\right)  .\label{line}%
\end{equation}
$g_{1}\left(  E/E_{P}\right)  $ and $g_{2}\left(  E/E_{P}\right)  $ are two
arbitrary functions which have the following property%
\begin{equation}
\lim_{E/E_{P}\rightarrow0}g_{1}\left(  E/E_{P}\right)  =1\qquad\text{and}%
\qquad\lim_{E/E_{P}\rightarrow0}g_{2}\left(  E/E_{P}\right)  =1.
\end{equation}
We expect the functions $g_{1}\left(  E/E_{P}\right)  $ and $g_{2}\left(
E/E_{P}\right)  $ modify the UV behavior in the same way as GUP and
Noncommutative geometry do, respectively. Following Ref.\cite{RGGM}, in
presence of Gravity's Rainbow, we find that Eq.$\left(  \ref{VEV}\right)  $
changes into%
\begin{equation}
\frac{g_{2}^{3}\left(  E/E_{P}\right)  }{\tilde{V}}\frac{\left\langle
\Psi\left\vert \int_{\Sigma}d^{3}x\tilde{\Lambda}_{\Sigma}\right\vert
\Psi\right\rangle }{\left\langle \Psi|\Psi\right\rangle }=-\frac{\Lambda
}{\kappa},\label{WDW3}%
\end{equation}
where%
\begin{equation}
\tilde{\Lambda}_{\Sigma}=\left(  2\kappa\right)  \frac{g_{1}^{2}\left(
E/E_{P}\right)  }{g_{2}^{3}\left(  E/E_{P}\right)  }\tilde{G}_{ijkl}\tilde
{\pi}^{ij}\tilde{\pi}^{kl}\mathcal{-}\frac{\sqrt{\tilde{g}}\tilde{R}}{\left(
2\kappa\right)  g_{2}\left(  E/E_{P}\right)  }\!{}\!.\label{LambdaR}%
\end{equation}
Of course, Eq.$\left(  \ref{WDW3}\right)  $ and Eq.$\left(  \ref{LambdaR}%
\right)  $ reduce to the ordinary Eqs.$\left(  \ref{WDW},\ref{VEV}\right)  $
and $\left(  \ref{LambdaSigma}\right)  $ when $E/E_{P}\rightarrow0$. By
repeating the procedure leading to Eq.$\left(  \ref{1loop}\right)  $, we find
that the TT tensor contribution of Eq.$\left(  \ref{WDW3}\right)  $ to the
total one loop energy density becomes%
\begin{equation}
\frac{\Lambda}{8\pi G}=-\frac{1}{3\pi^{2}}\sum_{i=1}^{2}\int_{E^{\ast}%
}^{+\infty}E_{i}g_{1}\left(  E/E_{P}\right)  g_{2}\left(  E/E_{P}\right)
\frac{d}{dE_{i}}\sqrt{\left(  \frac{E_{i}^{2}}{g_{2}^{2}\left(  E/E_{P}%
\right)  }-m_{i}^{2}\left(  r\right)  \right)  ^{3}}dE_{i},\label{LoverG}%
\end{equation}
where $E^{\ast}$ is the value which annihilates the argument of the root. In
the previous equation we have assumed that the effective mass does not depend
on the energy $E$. To further proceed, we choose a form of $g_{1}\left(
E/E_{P}\right)  $ and $g_{2}\left(  E/E_{P}\right)  $ which allows a
comparison with the results obtained with a Noncommutative geometry
computation expressed by Eq.$\left(  \ref{t1loop}\right)  $. We are thus led
to choose%
\begin{equation}
g_{1}\left(  E/E_{P}\right)  =\left(  1+\beta\frac{E}{E_{P}}\right)
\exp(-\alpha\frac{E^{2}}{E_{P}^{2}})\qquad\text{and}\qquad g_{2}\left(
E/E_{P}\right)  =1,\label{g1g22}%
\end{equation}
with $\alpha>0$ and $\beta\in%
%TCIMACRO{\U{211d} }%
%BeginExpansion
\mathbb{R}
%EndExpansion
$, because the pure \textquotedblleft\textit{Gaussian}\textquotedblright%
\ choice with $\beta=0$ can not give a positive induced cosmological
constant\footnote{See Ref.\cite{RGGM} for technical details.}. However this is
true when the effective masses satisfy relation $\left(  \ref{masses1}\right)
$. In case relation $\left(  \ref{masses2}\right)  $ holds the pure
\textquotedblleft\textit{Gaussian}\textquotedblright\ choice works for large
and small $x$, where $x=\sqrt{m_{0}^{2}\left(  r\right)  /E_{P}^{2}}$. The
final result is a vanishing induced cosmological constant in both asymptotic
r\'{e}gimes. It is interesting to note that Gravity's Rainbow has potential
effects on the photon propagation\cite{RGGM1}. Indeed, let us consider two
photons emitted at the same time $t=-t_{0}$ at $x_{dS}=0.$ The first photon be
a low energy photon ($E\ll E_{P})$ and the second one be a Planckian photon
($E\sim E_{P}).$ Both photons are assumed to be detected at a later time in
$\overline{x}_{dS}.$ We expect to detect the two photons with a time delay
$\Delta t$ given by the solution of the equation%
\begin{equation}
\overline{x}_{dS}^{E\ll E_{P}}(0)=\overline{x}_{dS}^{E\sim E_{P}}(\Delta t),
\end{equation}
that implies%
\begin{equation}
\Delta t\simeq g_{1}\left(  E\right)  \frac{e^{\sqrt{\Lambda_{eff}/3}t_{0}%
}-e^{\frac{\sqrt{\Lambda_{eff}/3}}{g_{1}(E)}t_{0}}}{\sqrt{\Lambda_{eff}/3}%
}\simeq\beta\frac{E}{E_{P}}t_{0}\left(  1+\sqrt{\Lambda_{eff}/3}t_{0}\right)
,\label{timed}%
\end{equation}
where we have used $\left(  \ref{g1g22}\right)  $ for the rainbow functions.

\bibliographystyle{aipproc}

\begin{thebibliography}{99}                                                                                               %
\bibitem {DeWitt}B. S. DeWitt, \textsl{Phys. Rev.} \textbf{160}, 1113 (1967).

\bibitem {Vilenkin}A. Vilenkin, \textsl{Phys. Rev.} \textbf{D 37}, 888 (1988).

\bibitem {Remo}R.Garattini, \textsl{J. Phys. A }\textbf{39}, 6393 (2006);
gr-qc/0510061. R. Garattini, \textsl{J.Phys.Conf.Ser}. \textbf{33}, 215
(2006); gr-qc/0510062.

\bibitem {CG:2007}S. Capozziello and R. Garattini, \textsl{Class.Quant.Grav.}
\textbf{24}, 1627 (2007); gr-qc/0702075.

\bibitem {Remo1:2004}R. Garattini, \textsl{TSPU Vestnik} \textbf{44}
\textbf{N7}, 72 (2004); gr-qc/0409016.

\bibitem {Remo1:2009}R. Garattini, \textit{The Cosmological constant and the
Wheeler-DeWitt Equation}, PoS CLAQG08 (2011) 012; arXiv:0910.1735 [gr-qc];

\bibitem {RG PN}R. Garattini and P. Nicolini, \textsl{Phys. Rev.} \textbf{D
83}, 064021 (2011); arXiv:1006.5418 [gr-qc].

\bibitem {MagSmo}J. Magueijo and L. Smolin, \textsl{Class. Quant. Grav.}
\textbf{21}, 1725 (2004) [arXiv:gr-qc/0305055].

\bibitem {RGGM}R.~Garattini and G.~Mandanici, \textsl{Phys. Rev.} \textbf{D
83}, 084021 (2011), arXiv:1102.3803 [gr-qc].

\bibitem {RGGM1}R.~Garattini and G.~Mandanici,\textit {Particle propagation and effective space-time in Gravity's Rainbow.}
To appear in Phys.Rev.D; e-Print: arXiv:1109.6563 [gr-qc].
\end{thebibliography}

\end{document}